# A deep learning framework for turbulence modeling using data assimilation and feature extraction

Atieh Alizadeh Moghaddam[1] & Amir Sadaghiyani[2]

## Abstract

*Turbulent problems in industrial applications are predominantly solved using Reynolds Averaged Navier Stokes (RANS) turbulence models. The accuracy of the RANS models is limited due to closure assumptions that induce uncertainty into the RANS modeling. We propose the use of deep learning algorithms via convolution neural networks along with data from direct numerical simulations to extract the optimal set of features that explain the evolution of turbulent flow statistics. Statistical tests are used to determine the correlation of these features with the variation in the quantities of interest that are to be predicted. These features are then used to develop improved partial differential equations that can replace classical Reynolds Averaged Navier Stokes models and show improvement in the accuracy of the predictions.*

## I. Introduction, motivation and objectives

Although today there are many precise methods for modeling problems in turbulent flow by computational fluid dynamic tools such as Large Eddy Simulation, Direct Numerical Simulation, etc., considering their high computational cost and difficulties, they cannot be used widely for industrial purposes. Currently the predominant tool for engineering problems with turbulent flows are Reynolds Averaged Navier Stokes (RANS) models. The accuracy of the RANS models is limited due to closure hypothesis that induce uncertainty into the RANS models. RANS model predictions are known to be unreliable for flows with strong pressure gradient, streamline curvature or flow separation. This includes both structural uncertainty and parametric uncertainty. These closure assumptions are used for predicting the Reynolds stresses by introducing functional dependency between mean flow properties and turbulent quantities that could be different in various turbulent problems. One of these assumptions is the turbulent viscosity hypothesis that equates the instantaneous Reynolds stresses to a function of the mean rate of strain with an isotropic eddy viscosity being the constant of proportionality. Another source is the gradient diffusion hypothesis that assumes that the transport of a turbulent quantity is along the mean gradient. Structural uncertainty in RANS models is a result of inadequacy in these closure hypotheses to represent the underlying physics in turbulent flows. Parametric uncertainties imply the uncertainty which is present in selection of calibration coefficients in turbulent flow. The coefficients in the model forms of RANS models are tuned to give good agreement in specific turbulent flows. This agreement may not hold true in other turbulent flows that were not used in the calibration of these coefficient values. The values of these coefficients depend on the flow being considered due to the limitations of the model form. This variability in the coefficient values leads to parametric uncertainty.

To study parametric uncertainty in RANS models, it is assumed that the model coefficients are uncertain and assign ranges or probability distributions for each of the model coefficients and quantify how this affects the quantity-of-interest. This approach also characterizes the sensitivity of the results to the choice of the coefficients. The methodology of Oliver & Moser [1,2] uses a stochastic extension to the Boussinesq hypothesis. This approach uses Bayesian inference and estimates the disagreement of the predictions of the RANS model with experimental data. Edeling et al [3,4] have used a similar methodology combining Bayesian inference with experimental data to infer the uncertainty in model coefficients of RANS models.

---

[1] Civil and Environmental Engineering (Tarbiat Modares University)
[2] Communication Systems Engineering (Shiraz University of Technology)

To study the structural uncertainty in RANS models, the form of the model is investigated for its limitations. A popular methodology to study structural uncertainty in RANS models is the perturbation method developed by Iaccarino and others [5,6]. This injects perturbations into the modeled Reynolds stresses towards the limiting states in the physically realizable ranges to study the sensitivity of the RANS predictions to the effect of small changes in the eigenvalues, eigenvectors of the Reynolds stress and the turbulent kinetic energy. This method has been applied successfully to complex turbulent flows in turbulent mixing [7], homogeneous turbulent flows [8], high-speed aircraft jets [9,10], turbulent flows through ducts [11], turbulent flow over hills [12], etc.

Recently there is a growing interest in turbulence modeling to use data driven approaches that use machine learning algorithms with the data and results of existing experiments, direct numerical simulation and large eddy simulation studies to improve RANS models. The Reynolds stresses are connected to parameters and features of the RANS model and the connection can be extracted by the use of machine learning techniques. These approach leads to more accurate results than results of traditional RANS modeling. Ling & Templeton [13] have used machine learning algorithms to identify the important parameters using data from DNS and LES studies. Mishra & Iaccarino [14] have used elastic nets to identify regions of large discrepancy in RANS model predictions. Ling et al [15] have developed a new deep neural network architecture to develop improved Reynolds stress models using data. Wang et al [16] have developed a new method to integrate knowledge from data to develop improved RANS models. This method is referred to as Physics Informed Machine Learning (PIML). This method learns the functional form of Reynolds stress discrepancy in RANS simulations based on available data. There are many challenges in using PIML based RANS models:

- The predicted Reynolds stresses field should be smooth and should converge.

- There are limited number of DNS models because this method is applicable only for low Reynolds numbers and simple geometries then we have limited data for training physics-based machine learning turbulent models.

- PIML models depend on some physics-based features of the turbulent problems. These features should be investigated further. There are a lot of features and we should investigate which features have main effect the model by DNS data, Features should be stand for all properties of the problem like geometry, Initial condition, boundary condition and temporal condition and their correlations. It worth mentioning that spatial correlation structures are crucial factors and they should be considered and it should be determined that how to properly incorporate physical constraints on the spatial correlations [16].

- These features should be investigated for every individual computational control volume in the flow field, they determine how we can model the Reynolds stresses in the RANS equation at each computational step of the numerical solution [17].

- Reynolds stresses could be modeled by a partial differential equation based on velocities,

pressure, first order gradient of velocities, higher order gradient and these gradients should be approximate based on existing mean values of velocity and pressure which are calculated by averaged Reynolds method we should use a high accuracy gradient reconstruction method. Incorporated parameters should be efficiently selected.

- The partial differential equation used for modeling the Reynolds stresses should be calculated according to the extracted feature of the problem and different features should lead to different PDE models.

## II. Methodology

In the recent past deep neural networks and deep learning tools have become very popular. Deep learning is a subset of machine learning approaches where input features are transformed through multiple layers of nonlinear interactions in a neural network with multiple hidden layers [18]. Deep neural networks are becoming popular owing to their capability to represent complex internal interactions of dynamical systems and achieve good predictions for wide ranges of applications like voice recognition [19], video classification [20], etc. Deep neural networks are well suited for applications to fluid dynamics and turbulent flows as their architecture is flexible and can be optimized for feature identification, recognition and selection tasks. Figure 1 shows the common architecture of a feedforward artificial neural network. The input layer, the output layer and the hidden layers are shown.

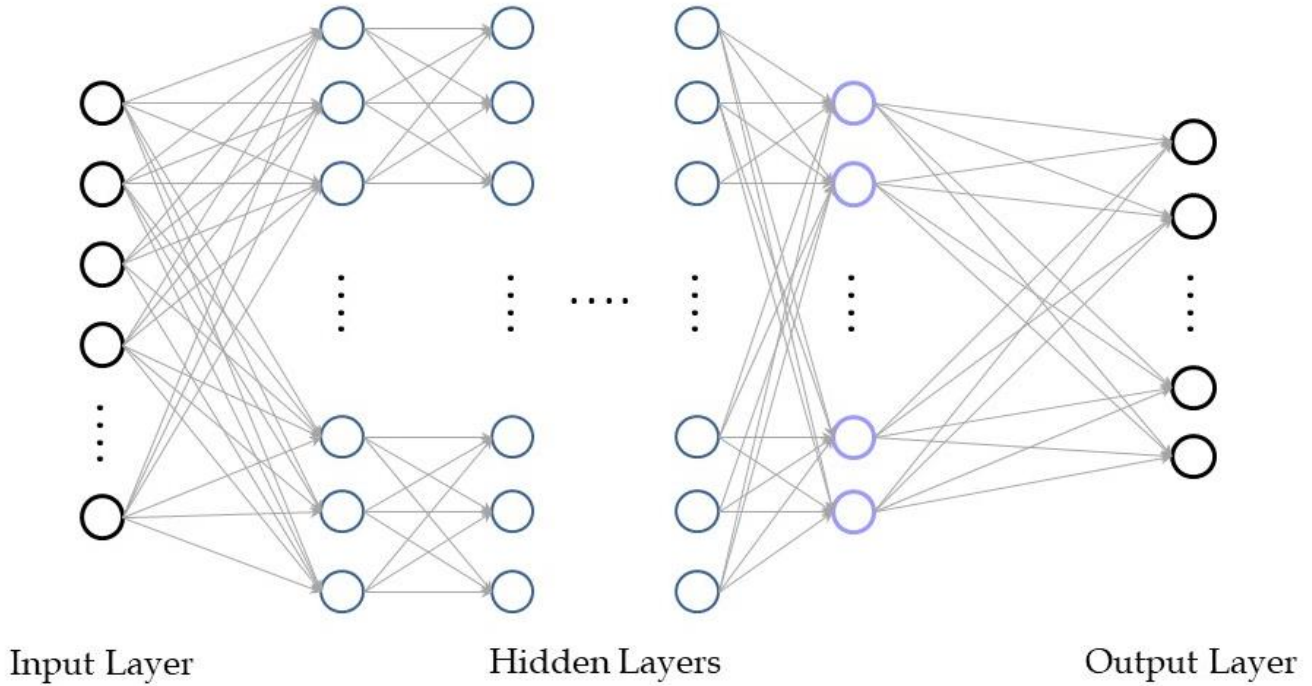

*Figure 1. General architecture of feedforward artificial neural network*

In this article we use a variant of neural networks called Convolution Neural Networks (CNNs) [21]. The depth and breadth of CNNs can be varied to control their power and computational cost. Compared to standard deep neural networks with similar size, CNNs have fewer connections and parameters. Because of this they are easier to train and their performance is almost comparable to standard ANNs.

To investigate the features we can use feature extraction method in deep learning approach for example by using convolution neural networks (CNN) on DNS data, these methods can extract parameters standing for features of problem (numeric and non-numeric characteristics). CNN uses filters to extract features and we can correlate these filters [22] and we can investigate the features and their relations in an integrate mode (geometrical, physical, etc.) we also can use image processing algorithms to determine geometrical features of the problem. These features are like a hidden layer in neural network and if we could extract them our neural network will be completed. We can compare different feature extraction methods.[17] The orthogonal partial least squares technique (OPLS) ,a successful strategy for feature extraction is introduced in [17] which can extract a few number of features of turbulent flows, by using the orthogonal partial least squares technique and project the input data onto the features. Feature extraction techniques, will extract all the features and the uncertainty can be controlled by used filters. [23] We propose to use an end-to-end neural network model, containing Multi-Scale Multi-modal Convolutional Neural Networks, which incorporates joint-feature extraction, selection and classification in a single framework. Multimodal machine learning is used to build models that can process and relate information from multiple modalities. Turbulence is a multimodal matter that is affected by different aspects of the problem [31]. We can design unsupervised kernel learning approach based on extracted features, inputs and outputs or we can use backpropagation methods to determine the convolutional kernels in hidden layers.[32,33]

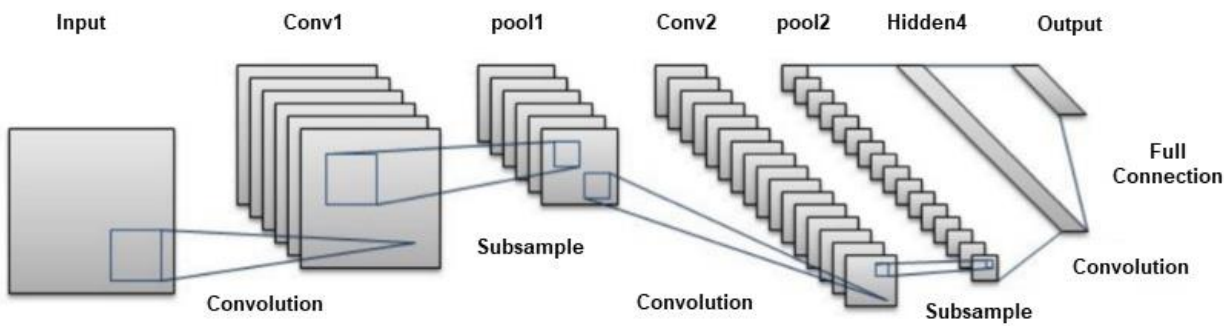

*Figure 2. General architecture of Convolution Neural Networks*

After investigating these features we can use a feature selection algorithm to discover main features, it happens based on the scores in various statistical tests for their correlation with the outcome variable, it reduces the CPU time. We name these selected features by $f_1$ to $f_M$. [24] These features can consider the spatial and temporal correlations of the variables [16].

Since the DNS data are limited we should use these data more efficient we can for example use a conformal mapping technique for transforming complex geometries to a simple geometry in Cartesian coordinates. Deep convolutional decoder/encoder networks are used for image segmentations and mapping inputs to appropriate outputs, so we can transform complex geometries and conditions to the standard geometries that are defined for the feature extraction convolutional neural network. The DNS data can be easily calculated for these standard geometries. [34, 35]

Extracted features guide us to determine the governing PDE for Reynolds stresses, we define $P_i$ as an indicator for velocities, pressure, gradients of the velocities (first order to higher order), characteristic locations (location vectors of the extreme points of velocity and pressure field or extreme of the gradients fields).

The final form of the partial differential equation is the inner product of the vector coefficients and the indicators given by

$$PDE = \sum_{i=1}^{n} \alpha_i P_i + \beta \qquad (1)$$

$\alpha_i$ are the coefficients and is a function of the selected features of the analysis. That is $\alpha_i = F(f_1, f_2, f_3 \ldots f_M)$.

$\beta$ is the smoothness term, this term guarantees the smoothness of the Reynolds stresses that are produced by the PDE. The value of Reynolds stress is unique in each point so the Reynolds stresses that are calculated in the interfaces of adjacent control volume with their different governing PDE's are equal then the smoothness factor can be extracted easily in each step. We can introduce another filters and modifiers to guarantee the smoothness and convergence of the average velocities and pressures based on extracted features.

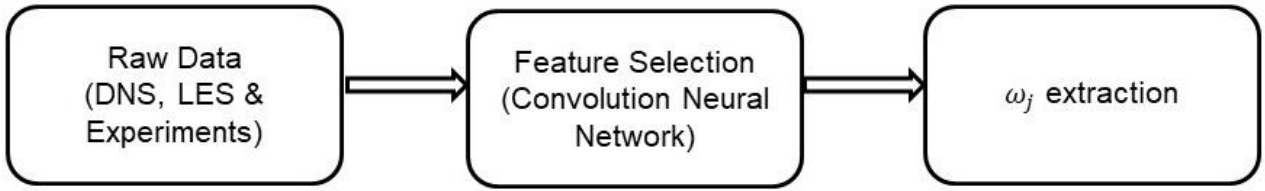

*Figure 3. Feature extraction from raw data using convolution neural networks*

The correct estimation of this functional dependence is important. We use a Gaussian radial basis function to determine F:

$$F = \sum_{j=1}^{N} \omega_j \exp(\varepsilon \sqrt{\sum_{i=0}^{n} (f_i - f_{i,j})^2})$$

Based on available DNS data we determine $\omega_i$ and for a new set of features of the new problem we can use $\omega_i$ and calculate the $F$ and the $\alpha$ coefficiens. The $\omega_i$ coefficients are constant for each feature in turbulence modelling frame work and only the features vary in different problems. Then we can predict Reynolds stresses modeling and by substituting in the RANS equations and discretization we will calculate the averaged velocity and pressure more accurate than traditional RANS models. We can modify the PDE to guarantee the numerical stability. The accuracy of the radial basis function interpolation and gradient constructions can be estimated [25], [26], [27].

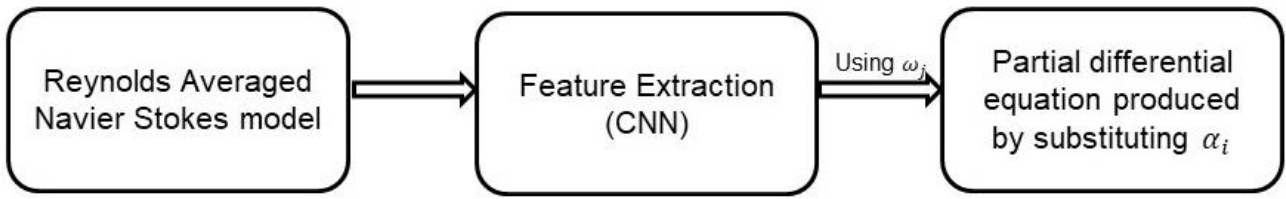

*Figure 4. Appending to classical RANS models using the extracted features*

## III. Outline

Using feature extraction method for turbulence problems is a new approach these features can encode the problem and contain all characteristics of the problem that has a role in turbulence modeling such as geometrical conditions, boundary conditions and temporal conditions and their integrated and joint effects. Deep learning architectures among which multi-modal convolution neural network processing jointly geometrical and physical modalities will be used for raw data (from direct numerical simulations, large eddy simulations, experiments and databases [28,29]) and they will establish a framework for turbulence modeling.

Extracting a partial differential equation based on velocities, pressure, locations of their extreme points, and fluid characteristics and, etc. This partial differential equation can be used instead of conventional closure hypothesis which induces high structural and parametric uncertainty to the problem. This method can leads to a smooth field of Reynolds stresses.

We consider all the correlations between extracted features as a new feature a multi-modality feature extraction will jointly investigate physical and geometrical features.

This method based on convolution neural networks can extract features from boundary condition and geometrical constraints. Temporal features can represent the history of the features. Extracted features will consider the effect of all boundary conditions on a point of flow field.

We propose a dividing technique to divide a complex geometry to some simple one and consider their effects on each other by some boundary condition, then by using a conformal mapping [30] each part of divided geometry can become simpler. By these techniques we wouldn't need DNS data of complex geometries.
Existing physical theories can be used as general ground hypothesis in deep learning architecture to compensate the lack of training data by providing prediction algorithms and extrapolation methods.

## IV. Conclusions

In this article we introduce a method to apply Convolution Neural Networks along with data from direct numerical simulations, large eddy simulations and experiments to extract optimal sets of features that explain the evolution of turbulent flow statistics. Statistical tests are also used to determine the correlation of these features with the variation in the quantities of interest that are to be predicted. These features are then used to develop improved partial differential equations that can replace classical Reynolds Averaged Navier Stokes models and show improvement in the accuracy of the predictions.